\documentclass[10pt,aps,amsmath,showpacs,amssymb,nofootinbib,pre,twocolumn]{revtex4}
\usepackage[dvips]{graphicx}
\usepackage[dvips,usenames]{color}
\usepackage{amsmath}

\begin{document}

\title{Slow, bursty dynamics as the consequence of quenched network topologies} 

\author{G\'eza \'Odor}
\affiliation{Research Center for Natural Sciences, 
Hungarian Academy of Sciences, MTA TTK MFA, 
P. O. Box 49, H-1525 Budapest, Hungary}

\date{\today}

\begin{abstract}

Bursty dynamics of agents is shown to appear at criticality or in extended 
Griffiths phases, even in case of Poisson processes. I provide numerical
evidence for power-law type of inter-communication time distributions 
by simulating the Contact Process and the Susceptible-Infected-Susceptible 
model. This observation suggests that in case of non-stationary bursty 
systems the observed non-poissonian behavior can emerge as the
consequence of an underlying hidden poissonian network process, which is 
either critical or exhibits strong rare-region effects. On contrary, in 
time varying networks rare-region effects do not cause deviation from the 
mean-field behavior and heterogeneity induced burstyness is absent.

\end{abstract}

\maketitle

\section{Introduction}

The dynamics of systems with general network communications has been
an interesting topic of various models and empirical observations
\cite{dorogovtsev07:_critic_phenom,barratbook}. In networks with 
large topological dimension defined as $N\propto r^d$, where $N$ 
is the number of nodes within the (chemical) distance $r$, the 
evolution is expected to be exponentially fast.
Generic, slow power-law type of dynamics are reported in 
\cite{Johnson,pv01a,pv04,vir,Chialvo,MM13}.
In social and neural networks the occurrence of generic slow dynamics was 
suggested to be the result of non-poissonian, bursty behavior of 
agents \cite{BA10} connected by small world networks 
\cite{vir,LLD13,KKBK12,KKK12}.  Times between contacts \cite{F2F} or 
communication \cite{K03,E04} between individuals was found to
deviate from a Poisson process, namely an intermittent switching
between periods of low activity and high activity, resulting in
fat-tailed inter-communication time distributions \cite{BA05}.

On the other hand arbitrarily large, rare-regions (RR), that can change 
their state exponentially slowly as the function of their sizes can cause 
so called Griffiths Phase (GP) \cite{Griffiths,Vojta}, in which
slow, non-universal, power-law dynamics occurs \cite{PhysRevLett.105.128701}.
It has been shown \cite{PhysRevLett.105.128701,odor:172,Juhasz:2011fk}
that GP-s can emerge as the consequence of purely topological disorder.
However, this has been found only in finite dimensional networks,
or in weighted tree-like networks for an extended time window
\cite{BAGPcikk,wbacikk,basiscikk}.

Griffiths singularities affect the dynamical behavior both below and above 
the transition point and can be best described via renormalization 
group methods in networks \cite{Monthus,KI11,JK13}. GP-s were
shown by optimal fluctuation theory and simulations of the Contact Process
(CP) \cite{harris74,liggett1985ips} on Erd\H os-R\'enyi (ER) \cite{ER} 
and on Generalized Small World (GSW) networks \cite{an,Juhasz3,Juhasz}.

The Susceptible-Infected-Susceptible (SIS) model \cite{SIS} is another 
fundamental system to describe simple epidemic (information) possessing 
binary site variables: infected/active or healthy/inactive. 
Infected sites propagate the epidemic (or active) all of their neighbors 
with rate $\lambda$ or recover (spontaneously deactivate) with rate $1$. 
SIS differs from the CP in which the branching rate is normalized by $k$, 
the number of outgoing edges of a vertex, thus it allows an analytic 
treatment, using symmetric matrices.
By decreasing the infection (communication) rate of the neighbors 
a continuous phase transition occurs at some $\lambda_c$ critical 
point from a steady state with finite activity density $\rho$ to an 
inactive one, with $\rho=0$ (see \cite{DickMar,rmp,odorbook}). 
The latter is also called absorbing, since no spontaneous activation 
of sites is allowed.

Very recently it has been proposed \cite{actdri} that many networks
can't be considered quenched ones, but evolve on the same time scale as
the dynamical process running on top of them. Activity driven network
models have been introduced, in which at a given time nodes possess only
a small number ($m=2$) of edges selected via a fixed, node-dependent 
activity potential $V_i$, which exhibits the probability distribution 
$F(V)\propto V^{-\gamma}$. Asymptotically the integrated link distribution 
is shown to be a scale-free (SF) network with $P(k)\propto k^{-\gamma}$ 
degree distribution \cite{actdri}.
In this work I investigate by extensive numerical simulations 
if rare-region effects and bursty dynamics could be observed in such 
networks with CP or Annihilating Random Walk (ARW) (see \cite{rmp}) 
processes running on them.

\section{Burstyness in the critical Contact Process }

\begin{figure}
\includegraphics[height=5.5cm]{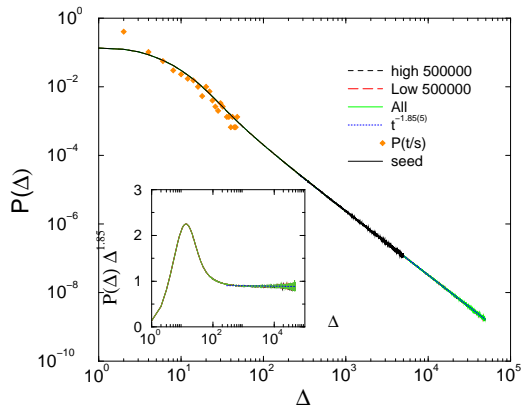}
\caption{\label{Pticp}(Color online) Inter-communication time distribution in the 1d
critical CP of size $N=10^5$. Full line denote histogramming from 
all times, dashed line from high times, long dashes from low times.
The dotted line shows a power-law fit for $t>200$ resulting in
$\propto t^{-1.85(5)}$. The solid thin line corresponds to runs from
seed initial conditions. Inset: the same data multiplied by the 
$t^{1.85}$ corresponding to the tail decay.}
\end{figure}

The $1d$ critical CP was simulated on rings of size $N=10^5$. 
The system was started from fully occupied state up to $t=10^6$ 
Monte Carlo steps (throughout this paper time is measured in MCs and 
shown to be unit-less on the figures). MCs are built up from full
sweeps of active sites. In one elementary MCs an active site is 
selected randomly and the activation is removed with probability 
$1-p=1/(1+\lambda)$, alternatively one of its randomly selected 
neighbor is activated with probability $p=\lambda/(1+\lambda)$.
The simulations were done around the critical point 
$\lambda_c=3.29785$ \cite{HH} of the CP. During
the simulations the times and the inter-communication times ($\Delta$) 
of neighbor activations of sites are calculated and histogrammed. 
Following the repetition of $\sim 200$ independent runs these 
timing data were analyzed and the probability distribution 
$P(\Delta)$ is calculated (see Fig.~\ref{Pticp}).

The systems during the runs are in non-stationary state, hence the 
average $\Delta$ should increase getting close to extinction, 
still $P(\Delta)$ for finite sizes exhibits a power-law tail, 
characterized by $P(\Delta)\sim \Delta^{-x}$ with the exponent
$x=1.85(5)$, obtained by least square fitting to the data.
To check if the non-stationary would cause a change in $P(\Delta)$ the 
histogramming was performed for the early ($t \le 5 \times 10^5$ MCs) 
and late ($t > 5 \times 10^5$ MCs) events separately. One cannot see 
any differences, all three $P(\Delta)$ distributions exhibit power-law
behavior.
On the other hand the $P(\Delta)$ distributions above or below 
$\lambda_c$ show exponential tails as expected.

The scaling behavior at the critical point can be derived by expressing
the inter-communication probability via the temporal auto-correlation function.
Infection events can happen if there is an infected-uninfected neighbor 
pair, a kink in the spin language: $n_i(t)=1$, at site $i$ and time $t$.
Using the two-time auto-correlation function $\Gamma(t,s)$ one can estimate
the probability of the subsequent infection events, separated by a
communication-free period $\Delta$ as:
\begin{equation}\label{deces}
P(\Delta) \simeq p^2 \Gamma(t,s) \prod_{j=1}^{j<\Delta} [(1-p)\Gamma(j,s)+(1-\Gamma(j,s))]
\simeq \Gamma(t,s)
\end{equation}
using the connected temporal correlator between times $s$ and $t$
\begin{equation}
\Gamma(t,s) = \langle n_i(t) n_i(s) \rangle  - \langle n_i(t) \rangle \langle n_i(s) \rangle
\end{equation}
Here $\langle\rangle$ denotes averaging for independent runs.
In this estimate the correlations among the inter-communication time
events are neglected, however this does not affect the asymptotic
behavior, because terms in the product are $\simeq O(1)$.

For 1d CP it is well known that this function exhibits an ageing 
behavior (see \cite{HDP,odorbook}), i.e. time translational invariance
is broken, but in the $t,s\to\infty$ limit the densities 
$\langle n_i(t) \rangle \to 0$ and the correlator scales as
\begin{equation}
\Gamma(t,s) \propto (t/s)^{-\theta} = (\Delta/s+1)^{-\theta}
\end{equation}
In case of 1d CP $\theta = 1.80(5)$ (see \cite{HDP,odorbook}). 
This is also true for the kink variables, which also follow the 
same universal scaling behavior, belonging to the directed 
percolation class \cite{odorbook}.
Strictly speaking due to the ageing behavior we have the scale
dependence $P(t/s) \sim \Delta^{-\theta}$ and indeed the simulations 
confirm this (see Fig.~\ref{Pticp}). 
Asymptotically one can find the same leading order contribution for 
$P(\Delta)$, coming from the smallest $s$ in the statistical average
and the tail behaviors agree with the auto-correlation function decay.

More generally, the site occupancy restriction condition of the CP is not
a necessary condition to find fat inter-communication tails. 
One can easily deduce, that the power-law tail of $\Gamma(t,s)$ of 
infections causes also fat-tails of the link-activation inter-communication 
times. This has been confirmed by the simulations.
Furthermore, simulation runs started from small activated seeds 
(see \cite{odorbook}) resulted in the same tail in $P(\Delta)$ again 
(see Fig.~\ref{Pticp}), only the distribution of activation times changes.
Contrary to the full initial condition case, where it decays as
$\sim \Delta^{-0.16(1)}$ it increases as $\sim \Delta^{0.33(1)}$ in the
case of seeds.

\section{Burstyness of the CP on generalized small-world networks}

In this section I show results obtained for the CP on certain GSW 
networks \cite{BB}. It has been shown that these system
exhibit extended GP regions, with non-universal, $\lambda$ dependent
power-law dynamics \cite{PhysRevLett.105.128701,Juhasz:2011fk}. 
The network generation starts with $N$ nodes on a ring.
All nearest neighbors are connected with Euclidean distance $l=1$ 
with probability $1$ and pairs with $l>1$ with a probability 
$p(l)=1-\exp (-\beta l^{-s})$.
For large distance this results in $p(l)\simeq \beta l^{-s}$.
Now I consider the cases: $s=2$ with $\beta = 0.1$ and $\beta =0.2$.

The inter-communication times of nodes were followed in networks with
$L=10^6$ nodes up to $t_{max}=10^6$ MCs as in case of the pure CP.
The number of independent samples at a given parameter, 
for which averaging was done, varied between $200$ and $1000$.
The $P(\Delta)$ distributions were determined for several $\lambda$-s 
in the GP of these networks. Invariance of $P(\Delta)$ with respect 
to the measuring time windows has been checked, similarly to 
the pure critical CP case. 

\begin{figure}
\includegraphics[height=5.5cm]{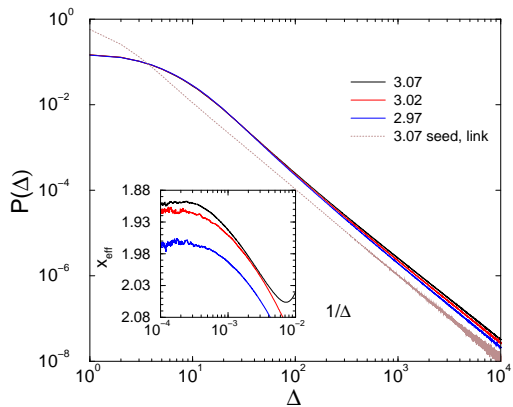}
\caption{\label{Pti1} (Color online) Inter-communication time distribution in the
GP of a GSW network with $\beta=0.1$ of size $N=10^6$ and 
$\lambda = 2.97, 3.02, 3.07$ (bottom to top solid curves).
The solid thin line corresponds to runs from seed initial conditions
measuring all activation attempts. 
Inset: Effective exponents defined as (\ref{xeff}) of the same data. }
\end{figure}

As Fig.~\ref{Pti1} shows power-law tails emerge again, with slightly $\lambda$ 
dependent slopes for $\beta=0.1$ at $\lambda=2.97, 3.02, 3.07$ within 
the GP region of the model. 
To see dependency on $\lambda$ and the corrections to scaling I applied the 
standard local slope analysis (see \cite{odorbook}) on the $P(\Delta)$ results. 
The effective exponent of $x$, which is the discretized, logarithmic 
derivative
\begin{equation}  \label{xeff}
x_\mathrm{eff}(t) = \frac {\ln P(\Delta) - \ln P(\Delta')} 
{\ln(\Delta) - \ln(\Delta')} \ ,
\end{equation}
where $\Delta/\Delta'=2$ difference was used here. 
As one can read-off from the inset of Fig.~\ref{Pti1}, at the critical 
point $\lambda_c=3.07(1)$, determined in \cite{Juhasz:2011fk},
$x$ tends to $1.90(1)$ asymptotically as $\Delta \to \infty$.
Below $\lambda_c$ the effective exponents converge to smaller values:
$x=1.92(1)$ at $\lambda=3.02$ and $x=1.96(1)$ at $\lambda=2.97$.
Corrections to the scaling are rather strong for $\Delta < 5000$,
but the effective exponents seem to saturate asymptotically.
Note, that as in case of the density decay study of this model
\cite{Juhasz:2011fk} logarithmic corrections were found in the GP.

As in case of the 1d CP the tail results are not affected by using
an active initial seed condition or by measuring the times between  
the communication attempts of sites. The combined effect of this two
modifications is shown on Fig.~\ref{Pti1} for $\lambda_c=3.07$. 

\begin{figure}
\includegraphics[height=5.5cm]{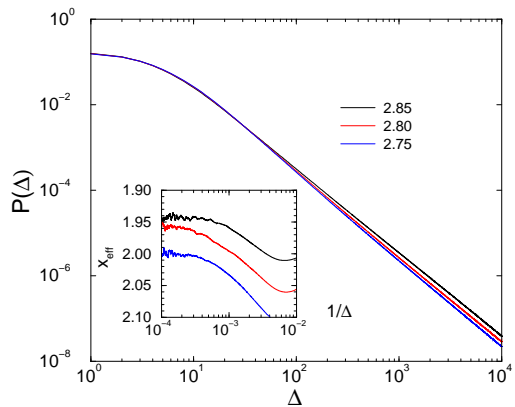}
\caption{\label{Pti2} (Color online) Inter-communication time distribution in the
GP of a GSW network with $\beta=0.2$ of size $N=10^6$ and 
$\lambda = 2.75, 2.8, 2.85$ (bottom to top curves). 
Inset: Effective exponents defined as (\ref{xeff}) of the same data.}
\end{figure}

For $\beta=0.2$ one finds somewhat different power-law tails
inside the GP (see Fig.~\ref{Pti2}). The local slope analysis suggests
$x=1.94(1)$ at $\lambda_c=2.85$, $x=1.96(1)$ at $\lambda=2.8$
and $x=1.99(1)$ at $\lambda=2.75$.
One can clearly see tail behaviors, characterized by increasing
$x$ exponents with $\beta$, in agreement with the fact that the 
addition of long edges to the network increases the topological 
dimension, thus the auto-correlation exponent, 
which is $\theta=4$ in the mean-field limit.

\section{Burstyness of the SIS model on ageing SF networks}

In this section I show SIS model results on ageing SF networks, 
where cutting the links among highly connected nodes results 
in finite topological dimension and GP behavior \cite{basiscikk}.
In the original Barab\'asi-Albert (BA)\cite{Barabasi:1999}
graph construction one starts from a single connected
node and add new links with the linear preferential rule.
In \cite{basiscikk} I investigated a generalized model, in which 
fraction of edges of the ageing nodes were removed from the BA
graph by a random, linear preferential rule. 
Consequently the edge distribution of the BA graph 
$P(k)\propto k^{-3}$ was cut off by an exponential 
factor for large $k$-s and quenched mean field theory suggested 
a GP behavior in agreement with the dynamical simulations.

\begin{figure}
\includegraphics[height=5.5cm]{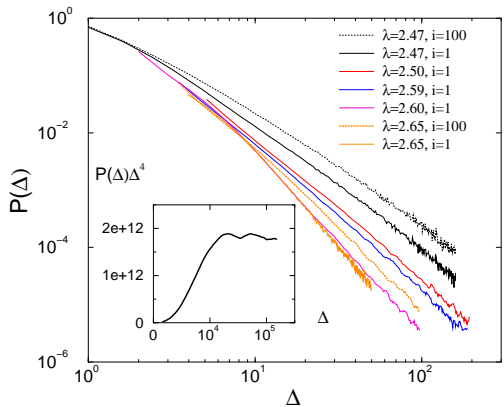}
\caption{\label{Ptig} (Color online) Inter-communication time distribution in the
GP of an ageing BA network of size $N=10^5$ for 
$\lambda=2.47, 2.5, 2.55, 2.59, 2.6, 2.65$ (top to bottom curves)
and measured at different ($i=1$ and $i=100$) sites.
Power-law fitting exponents for the tail behavior is shown in the 
text. Inset: $P(\Delta) \Delta^4 $ at the $\lambda_c$ of the CP defined
on the pure BA graph.}
\end{figure}

SIS model density simulations were run on systems with $N=10^5$
nodes in the formerly determined GP region $2.4 < \lambda < 2.7$ 
of the ageing BA graphs \cite{basiscikk}.
Occurrence of fat tail $P(\Delta)$ distributions can be seen on 
Fig.~\ref{Ptig}, but now even network site ($i$-)dependency emerges.
This is related to the fact that nodes are inhomogeneous:
the average number of edges decreases as $\langle k_i \rangle \sim i^{-1/2}$
by the BA network generation. Least squares error power-law fitting
for $\Delta > 20$ leads to $\lambda$ and $i$ dependent decay exponents.
For $\lambda = 2.65$ and $i=1$, (highest connectivity node) the
$P(\Delta)$ decay is characterized by the exponent $x=3.48(3)$, 
which is near to the mean-field value of the auto-correlation: 
$\theta = 4$ of the CP.
For less connected nodes ($i=100$) the decay is slower: 
$P(\Delta) \propto \Delta^{-2.96(3)}$, 
getting away from the mean-field value and coming closer to the 
one-dimensional auto-correlation exponent of CP. This agrees with
that our expectations, since for larger $i$-s the connectivity 
decreases and the system exhibits auto-correlations of lower 
dimensionality. By decreasing $\lambda$ in the GP, as shown in 
Fig.~\ref{Ptig} the following decreasing series of asymptotic 
tail exponents for $i=1$: 
$x = 3.48(3), 3.12(3), 2.64(2), 2.52(25), 2.42(2)$.
For $i=100$ at $\lambda=2.47$ the tail exponent is: $x=2.13(2)$.
Again, logarithmic corrections to the dynamic scaling can also expected
in the GP \cite{Vojta}.

To complete this study I also tested the critical point behavior 
of $P(\Delta)$ in case CP on the pure BA network (see \cite{BAGPcikk}) 
at $\lambda_c=1.21$. 
As the inset of Fig.~\ref{Ptig} shows the tail behavior tends to a 
power-law with $x=4$ for $\Delta > 20.000$ indeed.

\section{Dynamics of the CP and ARW on time varying networks} \label{sec:tv}

A simulation program has been created, with a fixed activity
potential $F(V)\propto V^{-\gamma}$ attached to vertexes, such that
two edges are connected to each node with that probability before
each 'sweep' of the network. One sweep (or Monte Carlo step) consists
of $N$ random CP updates of the network of $N$ nodes.
I followed $\rho(t)$ after a start from a fully occupied (infected)
state. The time is updated by one MCs after a full network sweep.
The simulations were run up to $t_{max}=2\times 10^5$ MCs on 
several sizes up to $N=10^7$ and repeated for $10^2-10^3$ independent 
randomly generated networks.

\begin{figure}
\includegraphics[height=5.5cm]{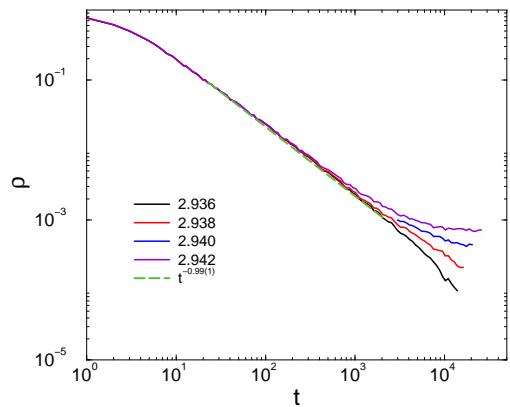}
\caption{\label{tw4cpl-3}(Color online)
Density decay of the Contact Process in time varying
networks of sizes $N = 10^7$ for $\lambda=2.936, 2.938, 2.940, 2.942$
(bottom to top). The dashed line shows a power-law fit to the  $\lambda=2.94$
(critical) curve. The activity potential decays with $\gamma=3$.}
\end{figure}

First $\gamma=3$ type of networks have been studied. The finite size effects
are strong, but for large sizes ($N \sim 10^7$) a phase transition seems to
appear with $\rho \propto 1/t$ decay, which agrees with the heterogeneous
mean-field prediction \cite{BAGPcikk} (see Fig.~\ref{tw4cpl-3}).
Similar results have been found for $\gamma=2.8$ networks.

\begin{figure}
\includegraphics[height=5.5cm]{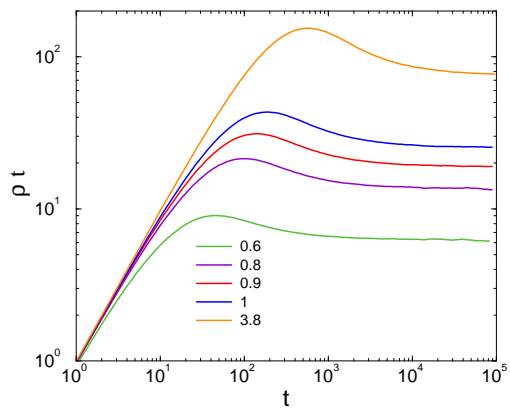}
\caption{\label{tar}(Color online) Density decay in the ARW in time varying
networks of sizes $N=10^7$ for $\gamma=0.6, 0.8, 0.9, 1, 3.8$
(bottom to top). The curves are the average of $10^3$ independent runs.}
\end{figure}

I have also tested the dynamical behavior of the Annihilating Random Walk 
(ARW) \cite{odorbook} in networks with activity potential
parameters: $\gamma=0.6, 0.8, 0.9, 1, 3.8$. 
The ARW model is a solvable model in homogeneous, Euclidean system, 
in which randomly selected particles hop to neighboring empty sites 
or annihilate with others on collision.
In the high dimensional, mean-field limit the density of particles decays
asymptotically as $t\propto 1/t$. As Fig.~\ref{tar} shows simulations
up to $t_{max}=10^5$ MCs with $N=10^7$ nodes result in the same 
asymptotic mean-field behavior following a long crossover time.
This suggests that slow, non-universal dynamics in activity driven
time varying networks do not exists, strengthening the hypothesis
\cite{odor:172} that quenched heterogeneity is a necessary condition for
observing rare-region effects.

Not so surprisingly burstyness does not occur in such time varying networks 
either, because the network rewiring process destroys the long-range 
dynamical correlations. Simulations result in an exponential tail 
$P(\Delta)$ distributions.

\section{Conclusions}

Observed burstyness in network systems is assumed to be the related 
of the internal, non-Poissonian behavior of agents or state variables. 
This has been explained by different multi-level or time-scheduling 
internal models.
In this paper I show an alternative route to this, being a natural 
consequence of correlated, complex behavior of the whole system.
In case of the critical, one-dimensional Contact Process fat-tailed 
inter-communication time distribution arises, related to the 
diverging auto-correlation function. 

Furthermore, the addition of long edges, which turns the network 
to GSWs with Griffiths Phases one can observe topology dependent, 
fat-tailed inter-communication time distributions.

I have also shown that in case of an ageing scale-free network, exhibiting
Griffiths Phase these power-law distributions depend also on the average
connectivity of nodes. The observed tail exponents vary in the range: 
$x=2 - 4$, which is smaller than the experimental values reported 
on human communication data-sets \cite{KKBK12,KKK12}. However, as the GSW
model example shows there exist networks, possessing smaller topological
dimensions, where $x < 2$. Furthermore, there are other models \cite{HDP}, 
like the bosonic contact process or the bosonic pair contact process, 
where the auto-correlation decays slower ($\theta=d/2$ for these 
unrestricted CPs \cite{BSH06}), thus $x$ could also be smaller on networks.

It is important to note that these systems are in a non-stationary state 
during the simulations, still the tail distributions are time invariant
and initial condition invariant.
Usually real systems are also in the non-stationary state as the 
consequence of various external conditions, circadian oscillations.
In case of regular networks the distributions are site invariant
as well.

Finally I have shown that both the Contact Process and Annihilating
Random Walks exhibit mean-field like dynamics on time varying, scale-free 
networks GP effects are absent and the distribution of inter-communication 
times is not bursty, but characterized by an exponential tail distribution.

These results suggest that bursty behavior can emerge as a collective
behavior in quenched network systems close to criticality or 
in extended GP like regions, suggesting a closer inspection of such
system. When real-world data confirms that sites exhibit inherent 
bursty behavior the superimpose of the two reason should emerge, 
possibly with the outcome of the more relevant one, which decays slower.

\section*{Acknowledgments}

I thank R. Juh\'asz, J. Kert\'esz, F. Igl\'oi and R. Pastor-Satorras 
for their useful comments. Support from the Hungarian research fund 
OTKA (Grant No. K109577), HPC-EUROPA2 pr.228398 and the European 
Social Fund through project FuturICT.hu (grant no.: 
TAMOP-4.2.2.C-11/1/KONV-2012-0013) is acknowledged.

\end{document}